\documentclass{aa}
\usepackage{natbib} 
\usepackage{graphicx}
\usepackage{txfonts}

\begin{document}

\title{The transition from quasar radio-loud to radio-quiet state\\
       in the framework of the black hole scalability hypothesis}

\titlerunning{The transition from quasar radio-loud to radio-quiet state}

\author{A. Marecki \and B. Swoboda}
\offprints{Andrzej Marecki \email{amr@astro.uni.torun.pl}}

\institute{Toru\'n Centre for Astronomy, N. Copernicus University,
           87-100 Toru\'n, Poland}

\date{Received 23 July 2010 / Accepted 23 September 2010}

\abstract
{}
{There are several lines of evidence that active galactic nuclei (AGN) can 
be regarded as scaled-up X-ray binaries (XRB). The timescales of the 
evolutionary phenomena in these two classes are proportional to the black 
hole (BH) masses. Consequently, unlike in the case of XRBs, the evolution of 
AGNs is too slow to be followed directly. What could be done, however, is to 
assign particular types of active galaxies to different evolutionary stages 
observable in XRBs. We studied such an assignment for three quasars with 
clear signatures of a recent transition from the radio-loud to the radio-quiet
state.}
{The quasars we investigated have large-scale radio lobes that are clearly 
asymmetric -- one lobe is of Fanaroff-Riley\,II type, while the other one is a
diffuse relic devoid of a hotspot. We suggest that the prime cause of the 
asymmetry of these radio sources is that the nuclei of their host galaxies 
currently produce no jets. To prove that, we observed them with 
milliarcsecond resolution to check if they are similar to those in 
radio-quiet quasars.}
{The observations carried out with the EVN revealed that the nuclei of the 
quasars under investigation are not of a core-jet type that is characteristic
for radio-loud, lobe-dominated quasars. It follows that the lobes are no longer 
fuelled and that the apparent asymmetry results from the orientation, which
causes a time lag of the order of 10$^6$ years between their images: the lobe 
perceived as a relic is nearer than the lobe with a hotspot and so it is 
observed in a later stage of the decay.}
{The three AGNs under investigation were radio-loud earlier, but now they have 
switched to the radio-quiet state. In the framework of the XRB/AGN 
unification, the above means that they have left the very high state and 
have moved now to the high/soft state. If this scenario is correct it poses 
a challenge to the so-called spin paradigm. While a radio-loud AGN must have
a spinning BH in its centre, the BH of a radio-quiet AGN does not necessarily
have low spin; AGNs with high-spin BHs, like those we deal with here, may
become radio-quiet.}


\maketitle


\section{Introduction}

In mathematical terms, black holes (BHs) are very simple objects. 
In principle, only two parameters, mass and spin, are sufficient
to describe them fully. Consequently, various similarities and analogies
between different classes of astronomical objects containing a BH in their
centres seem to be likely. Masses of BHs encountered in X-ray 
binaries (XRBs) on the one extreme and active galactic nuclei (AGNs) on the 
other span over eight orders of magnitude or more, and it is of 
interest whether any scaling laws connecting objects of otherwise quite 
different types like XRBs and AGNs exist -- see e.g. \citet{Fender2007} for 
a review. For the first time, a suggestion that this scaling actually takes 
place was made by \citet{SS1976}. A corollary of a possible unification of 
objects with accreting BHs based on their scalability is of great importance
because it paves the way for the extrapolation of some observable phenomena 
in one domain of BHs, namely stellar-mass BHs, to the other, i.e. to the 
domain of supermassive BHs (SMBHs) encountered in AGNs. Given that in 
particular characteristic timescales should be proportional to the mass, 
slow processes that are ongoing in AGNs could be directly followed in XRBs.

The first spectacular manifestation of putative BH scalability was 
discovered by \citet{Mirabel1992} who observed the X-ray source 
\object{1E\,1740.7$-$2942} with the Very Large Array (VLA) and found its
radio structure to be a double-sided jet straddling a compact and 
variable core. A striking similarity between the appearance of this 
phenomenon and that of radio-loud quasars (RLQs) led to coining the term 
``microquasar'' for jet-producing XRBs \citep{Mirabel1992}. However, this 
was \object{GRS\,1915+105} -- a highly variable XRB containing a BH of 
$\sim$15\,$M_{\sun}$ -- which became an archetypal microquasar, see the 
review by \citet{FB_ARAA}. It was discovered on 1992 August 15 in the
X-ray domain \citep{GRS1915_Xray} and a few months later in radio 
\citep{GRS1915_radio}. The most exciting property of this object was revealed
when \citet{MR1994} resolved its radio structure down to the fraction of an
arcsecond with the VLA and detected several components moving either toward
the east or west away from the core of GRS\,1915+105 with apparently
superluminal velocities. By the time of the announcement of that ground-breaking 
discovery, apparent superluminal motion (SLM) was already well known, but this
phenomenon was pertinent entirely to AGNs. Thus, finding the SLM in the vicinity
of a Galactic BH not only rendered the term ``microquasar'' most appropriate
for GRS\,1915+105 but also provided hard proof for the existence of an XRB
vs. AGN analogy.

This analogy was further confirmed by \citet{Marscher2002}, who monitored 
X-ray and radio emission of AGN in \object{3C120} radio galaxy and found 
dips in the X-ray emission that were followed by ejections of bright 
superluminal knots in the radio jet, just like in microquasars where dips in 
the X-ray emission precede radio flares. What was more, the mean time 
between the X-ray dips appeared to scale roughly with the BH mass. An extension 
of that monitoring programme carried out by the same group over five years 
(2002-2007) included observations at X-ray (2-10\,keV), optical (R and V 
bands), and radio (14.5 and 37\,GHz), as well as imaging with the Very Long 
Baseline Array (VLBA) at 43\,GHz \citep{Chatterjee2009}. The results of this 
comprehensive investigation were compelling enough to formulate a paradigm 
that XRBs and both radio-loud and radio-quiet active galactic nuclei were 
fundamentally similar systems, with characteristic time and size scales 
linearly proportional to the mass of the central BH. Also, the jet 
generation mechanism in microquasars and quasars seems to be the same, 
following the work of \citet{Tuerler2004}, who showed that the variability 
pattern of GRS\,1915+105 can be reproduced by the standard shock model used 
for extragalactic jets, e.g. that in \object{3C273}. Hence, the physical 
nature of relativistic jets does not depend on the mass of the BH and consequently
the study of jets in both quasars and microquasars is complementary. In line 
with the above results, \citet{McHardy2006} showed that characteristic 
timescales in the X-ray variations both in AGNs and Galactic BHs could be 
physically linked to an extent that it appeared that the accretion process was 
exactly the same for small and large BHs. Therefore, over a range 
$\sim$10$^8$ in mass and $\sim$10$^3$ in accretion rate, SMBHs can be 
regarded as scaled-up Galactic BHs.

However, as far as the scope of this work is concerned, perhaps the most 
interesting aspect of the XRB/AGN analogy was pointed to by 
\citet{Nipoti2005}. According to them, the well-known radio-loud (RL) vs. 
radio-quiet (RQ) dichotomy in quasars was reflected in certain states of 
radio/X-ray emission observable in microquasars: the fraction of observable 
RLQs -- $f_Q$\,$\approx$\,0.08 -- is compatible with the fraction of the 
time that microquasars remain in the so-called flaring state. Moreover, the 
authors posited that QSOs could switch between the RL and RQ states, but 
spend much more time in the latter one. This conjecture is testable 
observationally and the most straightforward way to show that it is correct 
is to find objects that are currently in a transition between RL and RQ states 
or are at least shortly after such a transition. At 
first sight, this seems hardly possible because the timescales of the 
evolutionary processes in AGNs are too extended for these phenomena to be 
directly observed. Yet we show here that these tests are possible and
present the results we obtained for three QSOs with
signatures of a recent transition from the RL to the RQ state. Based on our
results, we link this transition with its XRB analog.

\section{The method}

The well known upper limit for the spectral ages of double radio sources is of
the order of $\sim$10$^8$\,yr \citep{AL1987, Liu1992} and the upper 
limit for their linear sizes is $\sim$4\,Mpc, see e.g. \citet{Machalski2008}.
These two limits constitute compelling evidence that the energy transport 
from the RL AGN to its lobes cuts off eventually, and so the radio source 
fades out. The spectrum of a switched-off source becomes very steep because of 
radiation and expansion losses, and its lobes gradually disperse. According 
to \citet{KG1994}, the coasting phase of the lobes of a source that is no 
longer fuelled by its AGN can be very long: up to $10^8$\,yr. But
hotspots fade out much sooner; their lifetimes are roughly $7\times 
10^4$\,yr \citep{Kaiser2000}. Hence, the lack of well defined hotspots is an 
early signature of a lobes' decay.

There is yet another important observable phenomenon resulting from
cessation of the activity in the radio domain:
the lobe asymmetry, which, as pointed out by \citet{Marecki2006}, is a possible
signature of a radio activity switch-off. We mean here by ``asymmetry''
that one lobe is diffuse and devoid of a hotspot, while 
the other resembles a Fanaroff-Riley \citep{FR74} type-II (FR\,II) lobe 
quite well. In the extreme case, not only the hotspot but the whole lobe 
disappears. We interpret this as follows. If a double structure of a radio 
source does not lie in the sky plane (but also is not beamed towards us) the 
light-travel time plays a role: the epoch in which we observe the far-side 
lobe is significantly earlier than the epoch of the near-side lobe. For 
example, if the angle between the line of sight and the source axis is 
45$\degr$, the time lag is equal to the projected linear size divided by $c$ 
and attains a magnitude of the order of $\sim$10$^6$\,years for a typical 
large-scale, i.e. $\sim$10$^5$-pc sized radio source. This is definitely 
long enough for a hotspot to disappear as well for a lobe to disperse 
considerably. The above means that the lobe asymmetry is a very valuable 
signature of a recent cessation of RL state or -- to be more exact -- either 
a switch-off of the {\em whole} activity in the nucleus of the galaxy or at 
least of a transition from RL to RQ state. To prove that the latter 
scenario is correct, one has to show radio sources with asymmetric, i.e. one 
``still alive'' and one ``dying'' lobe straddling a QSO. This way, one 
proves that the host galaxy is still active but its nucleus no 
longer produces
jets, and so it has entered an evolutionary track leading to the fully RQ state. 
The presence 
of broad emission lines, which are best observed in QSOs, is essential to 
ascertain that in accordance with the AGN unification scheme 
\citep{Antonucci1993, UP1995}, the AGN axis makes a moderate,
i.e. neither small nor close to right,
angle with the line of sight so that the time lag between the lobes' 
images indeed takes place, and {\em this} is the most plausible and natural 
explanation of their apparent asymmetry.
Using QSOs for this purpose is even more justified
by the seminal work of \citet{Barthel1989}, who assert that the
axis of a QSO makes an angle of less than 44$\fdg$4 with the line of sight.
Therefore, if a transition from RL to RQ state occurs in a QSO 
with a large-scale, double-lobed structure,
it must result in this asymmetry. To make the proof complete, one 
should finally show that the high-resolution radio image of the object's 
core has the characteristics of those observed in ``true'' RQ quasars (RQQ)
known from 
the literature. Below we report how we carried out the above-sketched proof.

\begin{figure*}
\centering
\includegraphics[width=6.7cm]{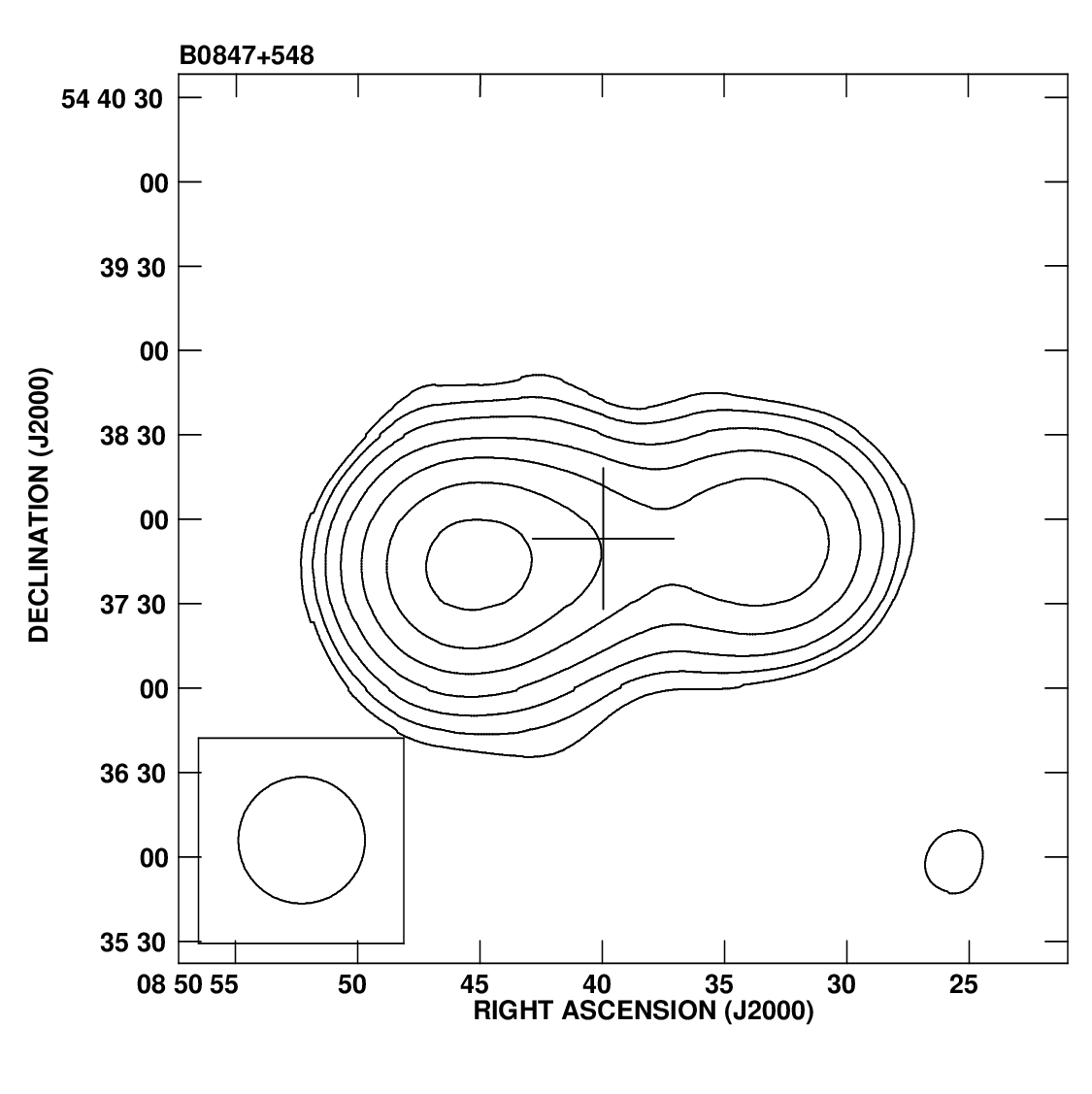}
\includegraphics[width=6.7cm]{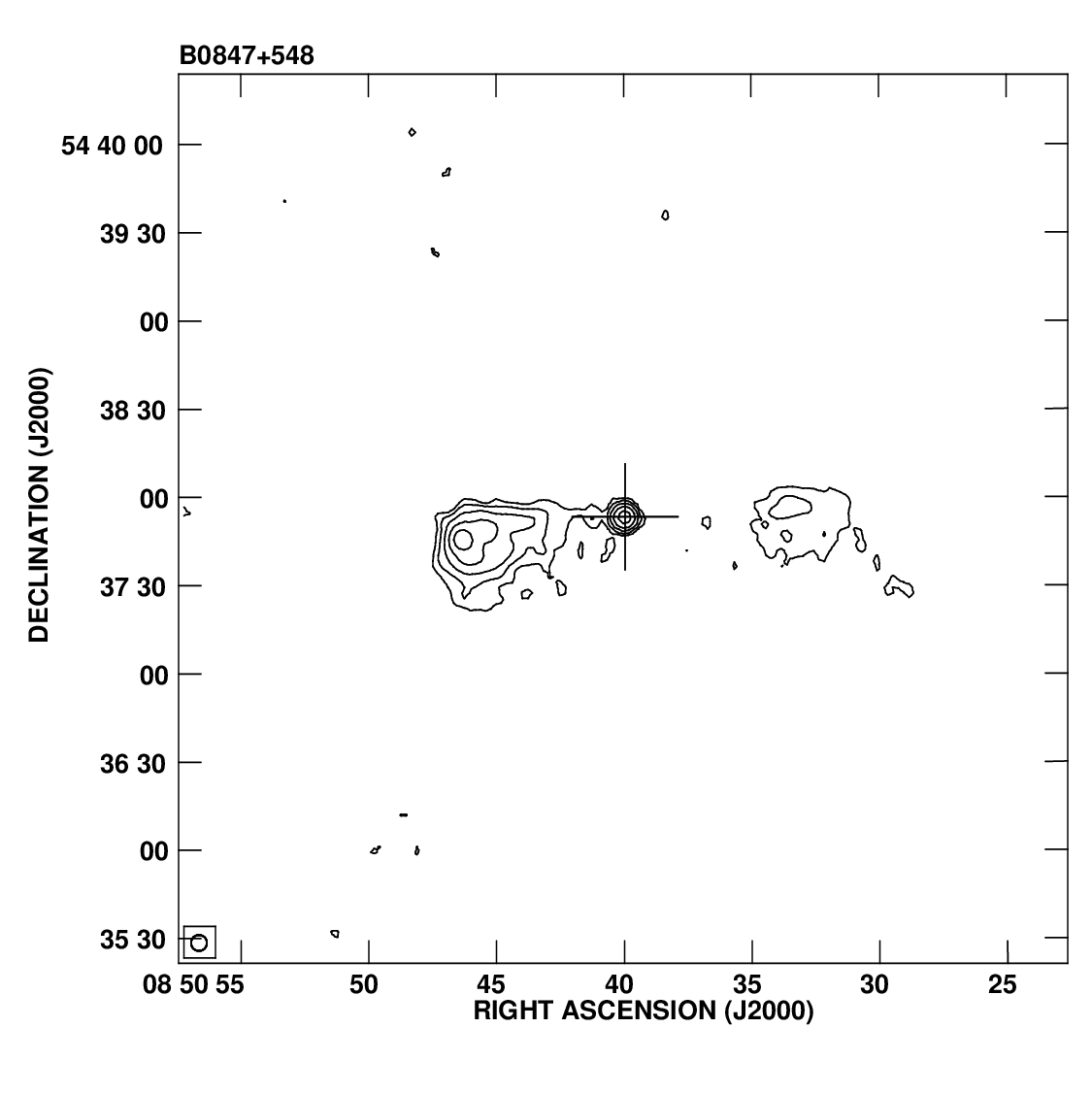}
\includegraphics[width=6.7cm]{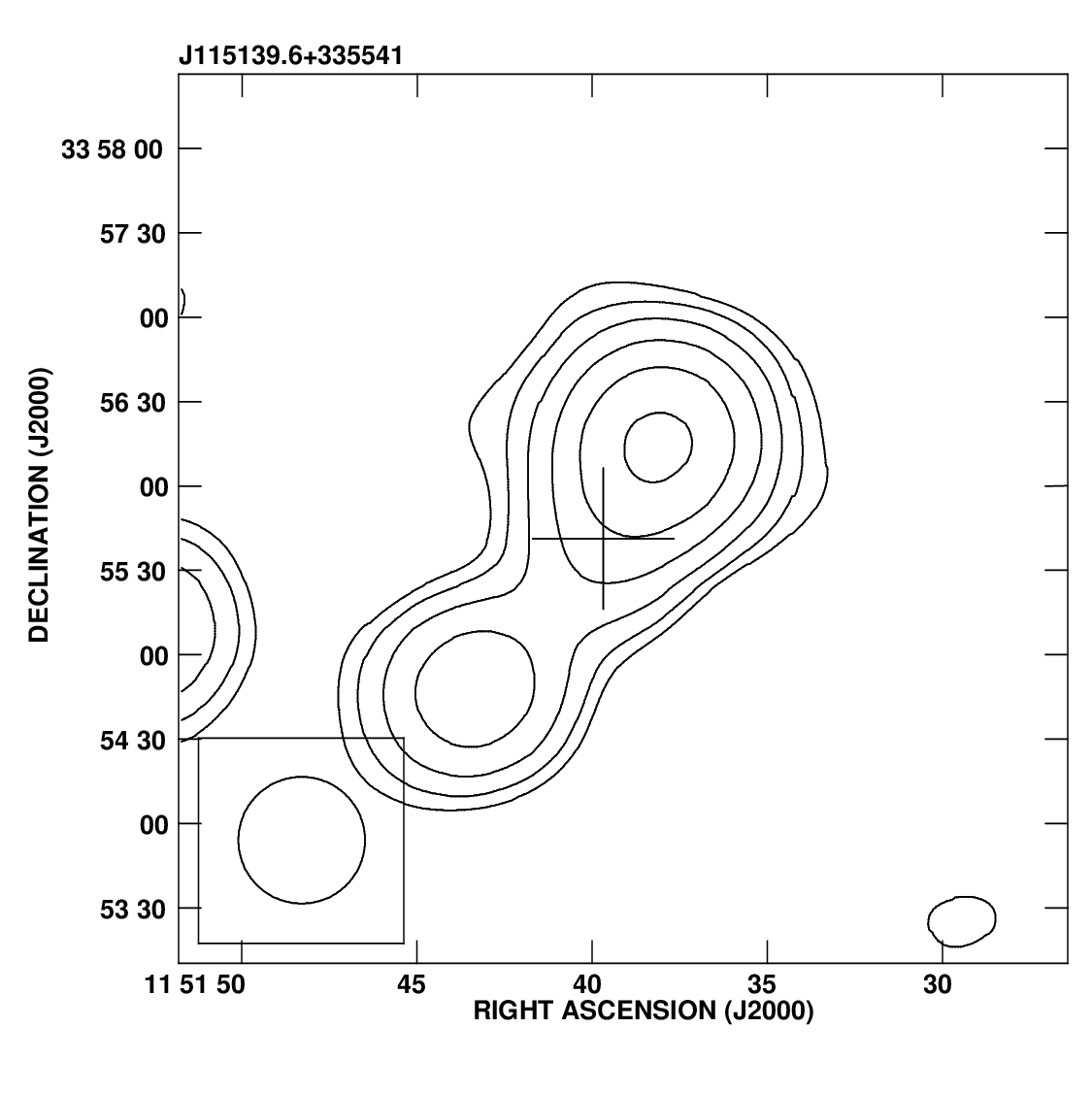}
\includegraphics[width=6.7cm]{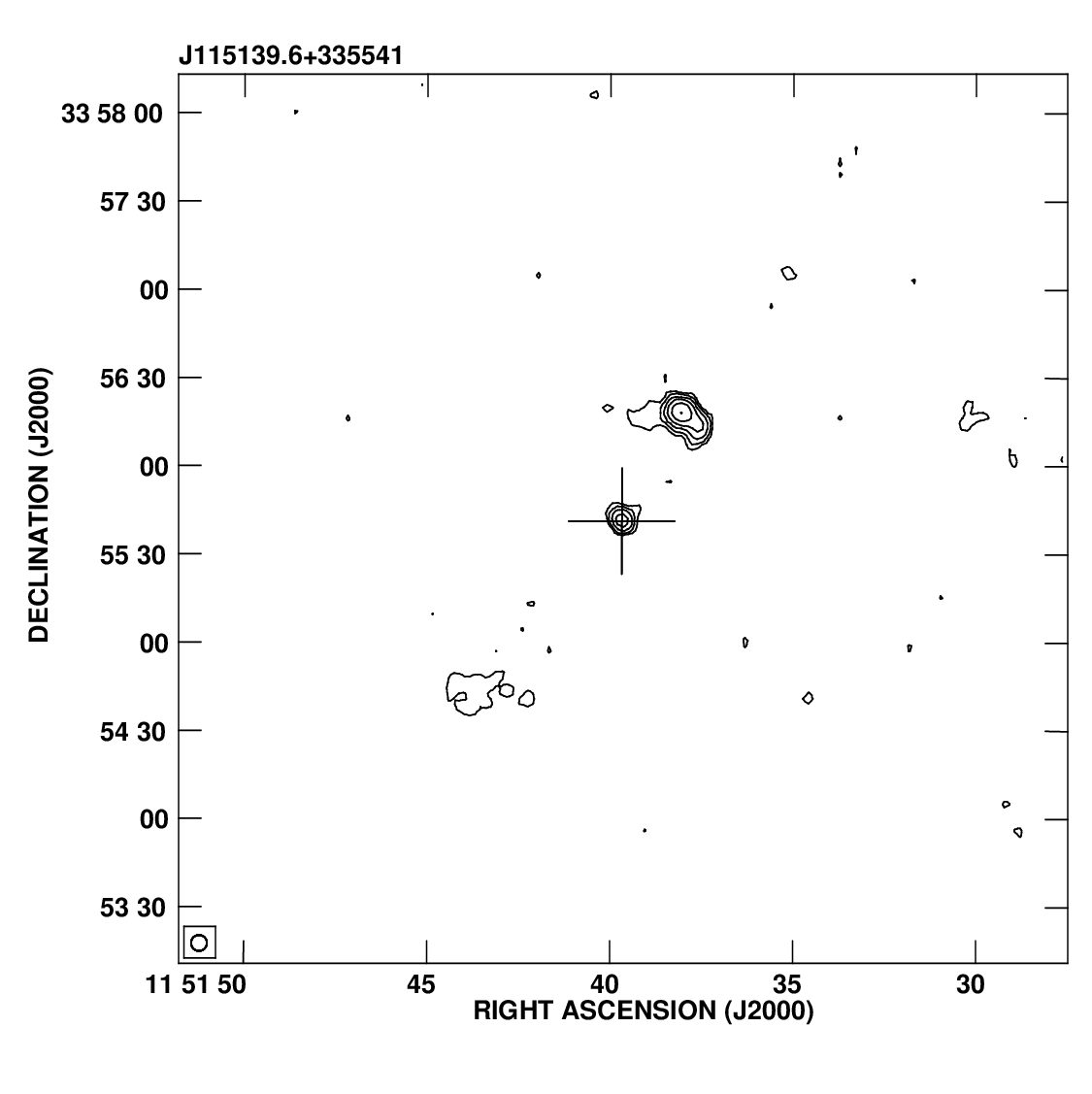}
\includegraphics[width=6.7cm]{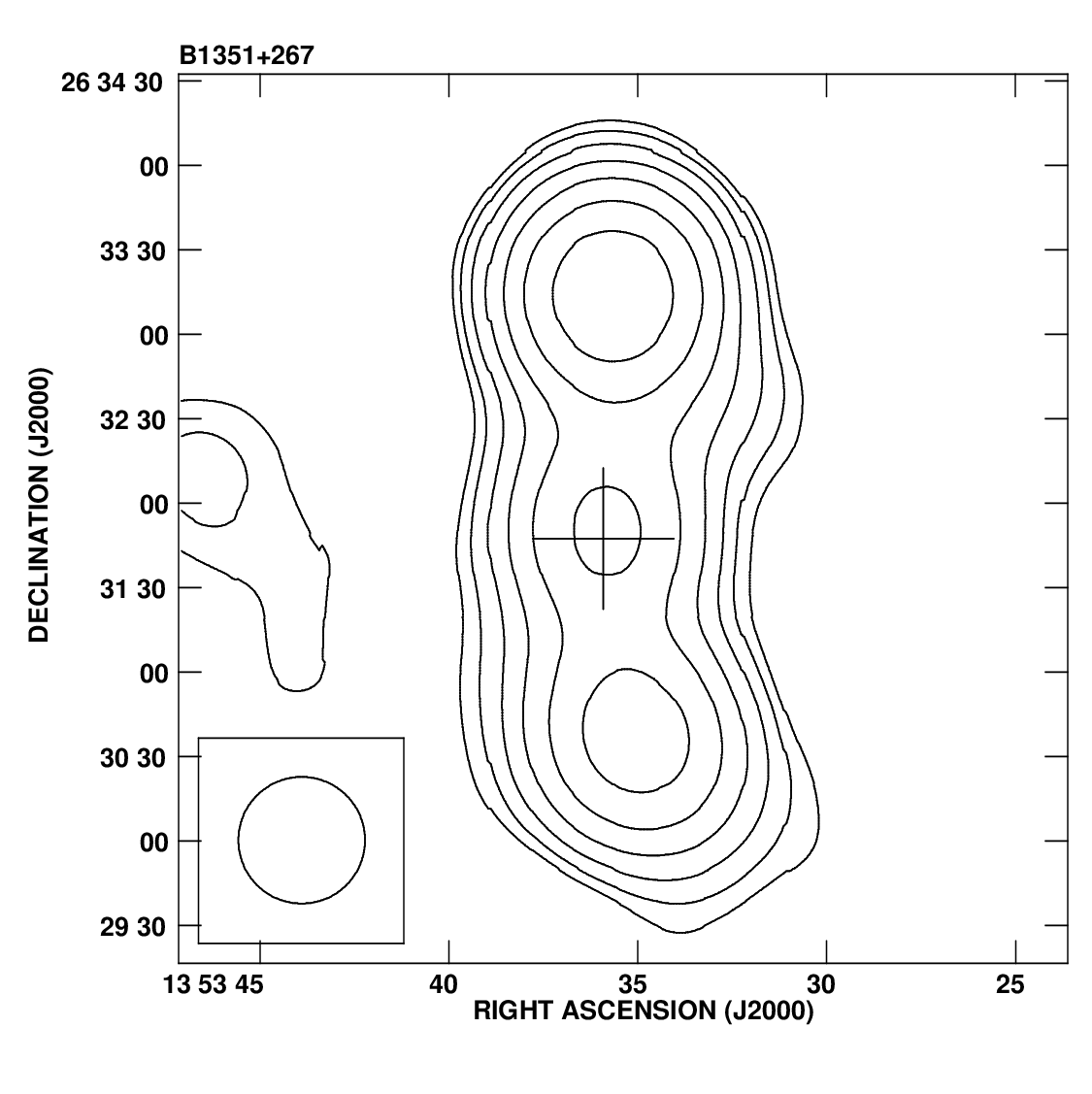}
\includegraphics[width=6.7cm]{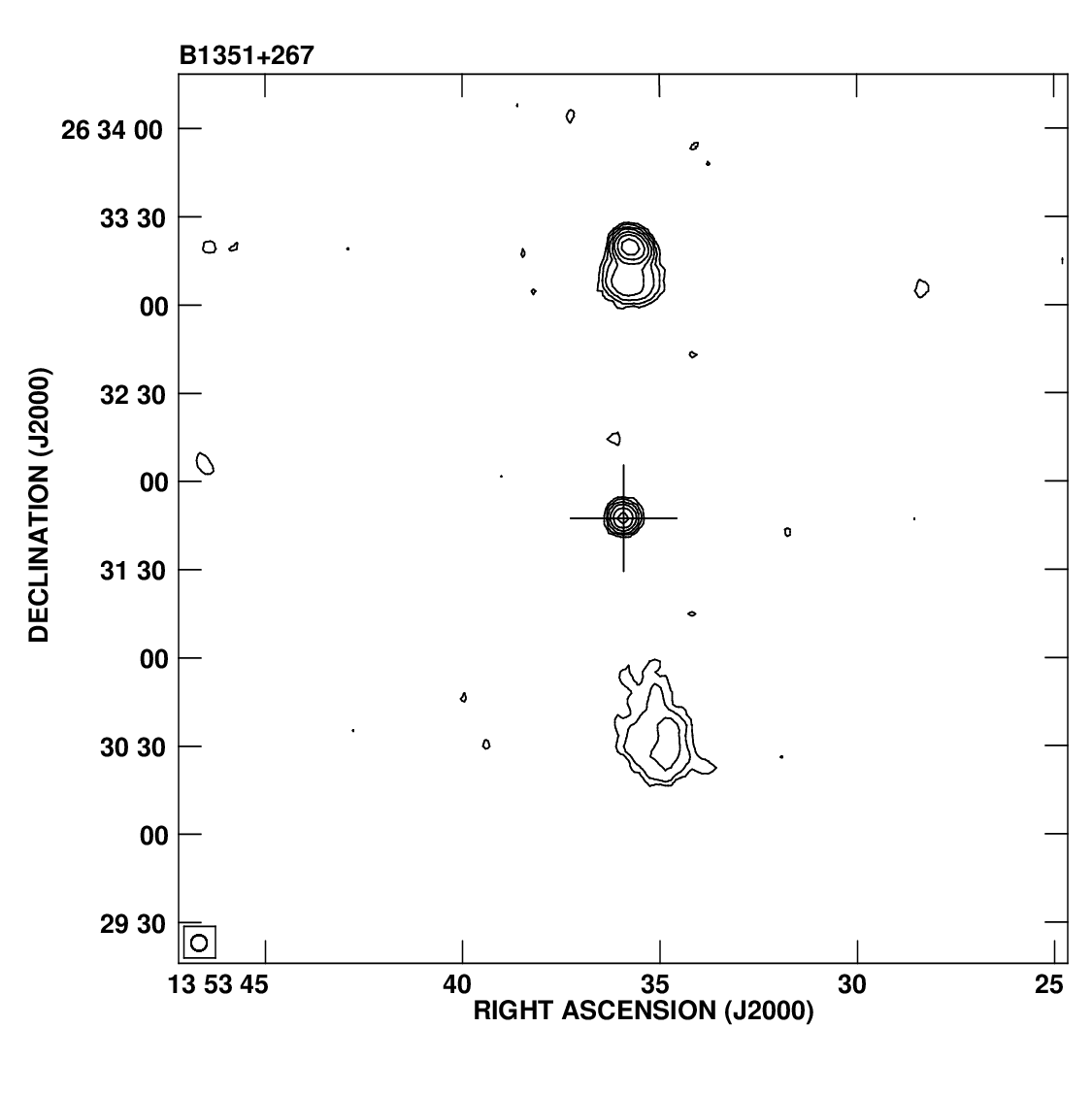}
\caption{NVSS (left column) and FIRST (right column) images of three quasars with
asymmetric lobes. Contours are increased by a factor of 2; the first contour level
corresponds to 1mJy/beam for NVSS images and 0.5mJy/beam for FIRST images.
Crosses indicate the positions of optical objects.}
\label{fig:NVSS_FIRST}
\end{figure*}

As a first step, we constructed a genuine sample of AGNs that have just 
switched off in the radio domain.
The procedure was as follows. We selected double radio sources 
as seen in {\it NRAO/VLA Sky Survey} (NVSS) \citep{Condon1998}\footnote{Website:
{\tt http://www.cv.nrao.edu/nvss}} with optical identifications 
located between the two putative lobes. To this end, we used the catalogue 
compiled by \citet{FH2004}\footnote{Website: {\tt 
http://quasars.org/qorg-data.htm}}. For a typical large-scale double, NVSS 
carried out with the VLA in D-conf. is well suited to indicate the presence 
of the lobes regardless of whether they are relics or not. This way, the 
algorithm associating optically-selected AGNs with double-lobed radio 
sources used by Flesch \& Hardcastle works reliably even if a given lobe is 
a relic. From this point on, the selection process was based on the 
elementary principle of radio interferometry: diffuse objects were poorly 
imaged without short spacings. Specifically, we relied upon the assumption 
that VLA in B-conf. used during the {\it Faint Images of the Radio Sky at 
Twenty-Centimeters} survey (FIRST) \citep{White97}\footnote{Website:
{\tt http://sundog.stsci.edu}} failed to reproduce lobes well or even at 
all if they were diffuse. Therefore, targets selected form NVSS embedded 
within FIRST footprint were processed with an algorithm calculating the 
ratio of peak-to-integrated flux densities for FIRST catalogue items 
pertinent to each of the two NVSS lobes. When that ratio was very low for 
all the FIRST components within the given NVSS lobe, which meant that the 
lobe was featureless and so likely to be diffuse, it was regarded as a 
potential relic. Additionally, we found that in a number of cases the NVSS 
component had no FIRST equivalents and, obviously, we treated them as true 
positives.

\begin{table*}
\caption{Basic parameters of the targets}
\label{tab:basic}
\centering
\begin{tabular}{l c c c c r c c}
\hline
\hline
Object name & RA & Dec & $z$ & Projected &
Core flux\tablefootmark{a} & \multicolumn{2}{l} {Flux of the relic
lobe\tablefootmark{a}}\\
& \multicolumn{2}{c} {(J2000)} & & linear size & & Peak & Integrated\\
& & & & [kpc] & \multicolumn{1}{c}{[mJy]} & [mJy] & [mJy]\\
\hline
B0847+548 & 08 50 39.981 & +54 37 53.19 & 0.367\tablefootmark{b}
& 567 & 13.49 & 1.36 & 33.69\\ 
J115139.6+335541 & 11 51 39.677 & +33 55 41.74 & 0.851\tablefootmark{b}
& 898 & 6.58 & --- \tablefootmark{c} & --- \tablefootmark{c}\\
B1351+267 & 13 53 35.925 & +26 31 47.54 & 0.307\tablefootmark{d}
& 752 & 23.63 & 2.92 & 39.55\\
\hline
\end{tabular}
\tablefoot{
\tablefoottext{a}{Extracted from FIRST catalogue.}
\tablefoottext{b}{Source: SDSS}
\tablefoottext{c}{Not itemised in FIRST catalogue.}
\tablefoottext{d}{Source: \citet{EH2004}}
}
\end{table*}

Application of the above automated procedure led to a selection of a few 
hundreds of potential targets. This approach was by no means perfect and it 
produced quite a number of false positives. To sieve them out, we downloaded 
all the FIRST images indicated by the algorithm and inspected them visually. 
Thirty-eight sources with one relic and one ``active'' lobe remained. They 
were mainly normal galaxies without a trace of a radio core, but several 
sources in our sample appeared to have cores anyway and a few of them were 
identified with QSOs. For three, the flux densities of the cores at 
1420\,MHz as seen in FIRST were more than 5\,mJy, which made these objects 
suitable targets for observations with the European VLBI Network (EVN).

\begin{figure}
\resizebox{\hsize}{!}{\includegraphics{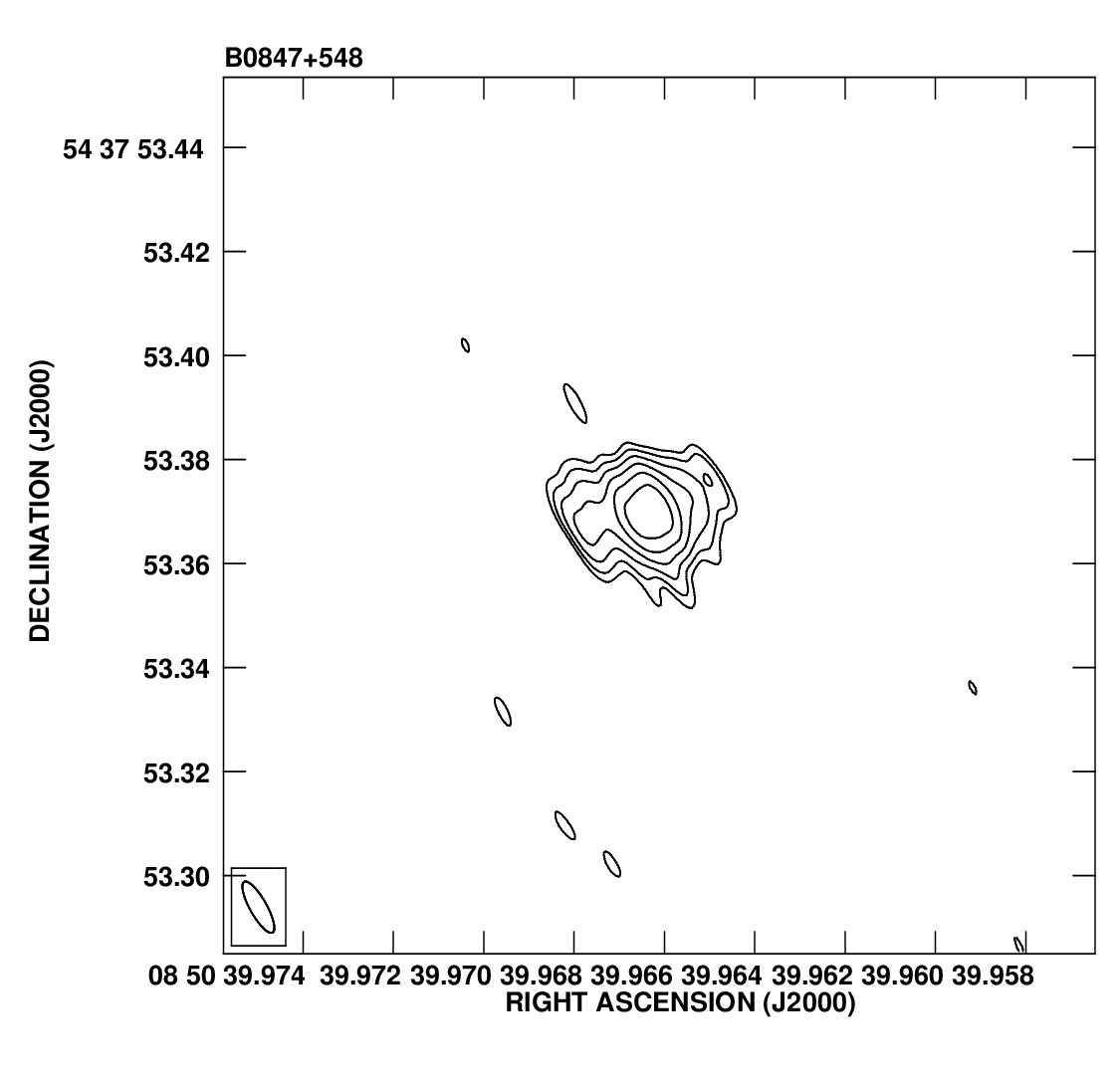}}
\caption{EVN image of the core of B0847+548. Contours are increased 
by a factor of 2; the first contour level corresponds to 5\,$\sigma$ level 
which is 52\,$\mu$Jy/beam. The peak flux density is 2.84\,mJy/beam. The beam 
size is $11.2 \times 3.2$\,milliarcseconds at the position angle of 
29$\degr$.}
\label{fig:B0847_EVN}
\end{figure}

\begin{figure}
\resizebox{\hsize}{!}{\includegraphics{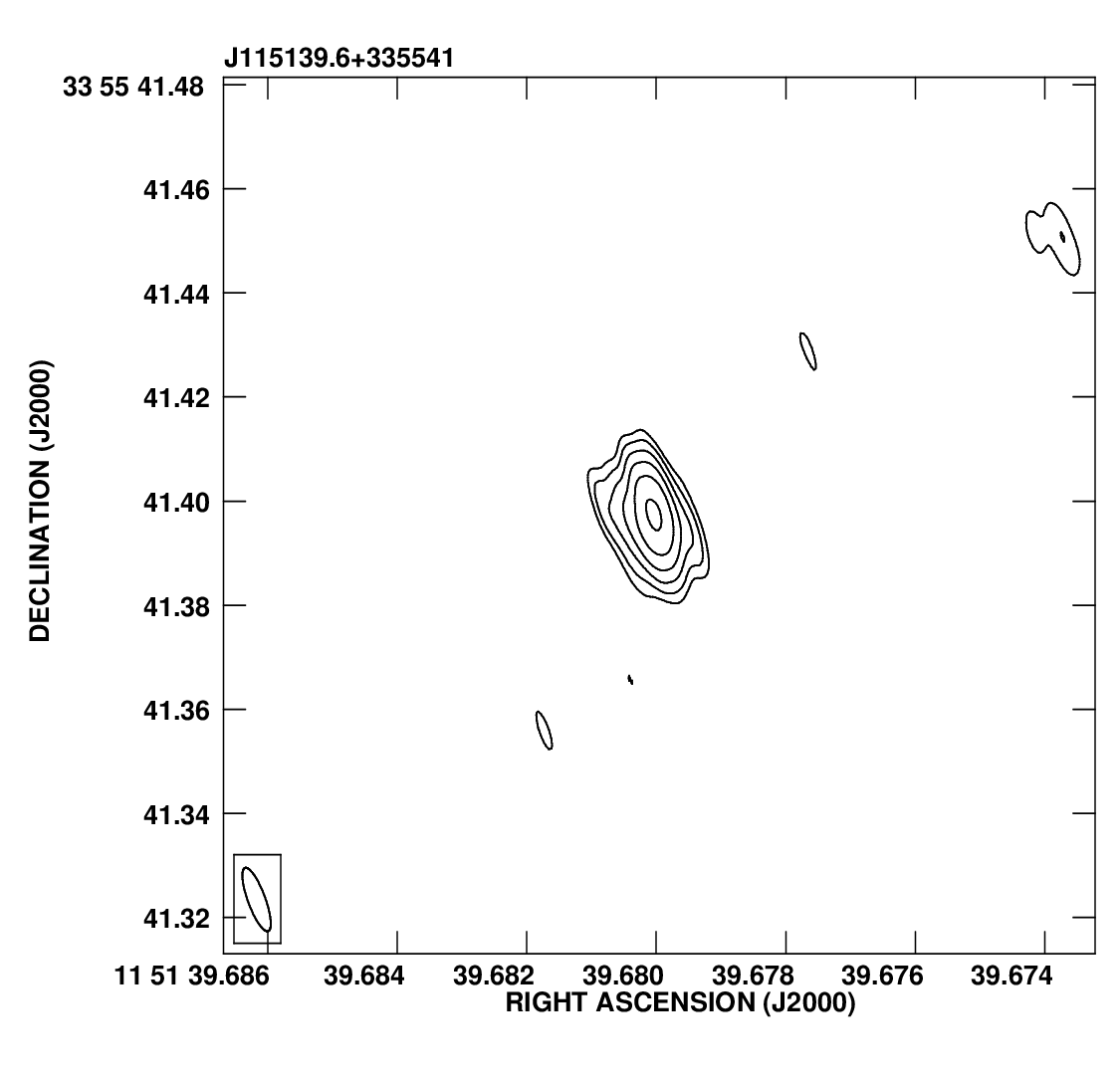}}
\caption{EVN image of the core of J115139.6+335541. Contours are increased 
by a factor of 2; the first contour level corresponds to 5\,$\sigma$ level 
which is 51\,$\mu$Jy/beam. The peak flux density is 1.85\,mJy/beam. The beam 
size is $12.9 \times 3.4$\,milliarcseconds at the position angle of 
20$\degr$.}
\label{fig:J1151_EVN}
\end{figure}

\section{Observational data}

\subsection{Radio domain}

The basic parameters of the radio sources selected for VLBI observations are 
given in Table~\ref{tab:basic} and their NVSS and FIRST maps are shown in 
Fig.~\ref{fig:NVSS_FIRST}. Based on the NVSS images, the radio structures of 
our targets are clearly double or triple with one lobe slightly weaker, but 
only the FIRST maps reveal the real cause of that weakness: lack of the 
hotspot and diffuse shapes. Quantitatively, the latter feature is well 
expressed by the ratio of peak-to-integrated flux density of the diffuse 
lobe as shown in the last two columns of Table~\ref{tab:basic}.

\begin{figure}
\resizebox{\hsize}{!}{\includegraphics{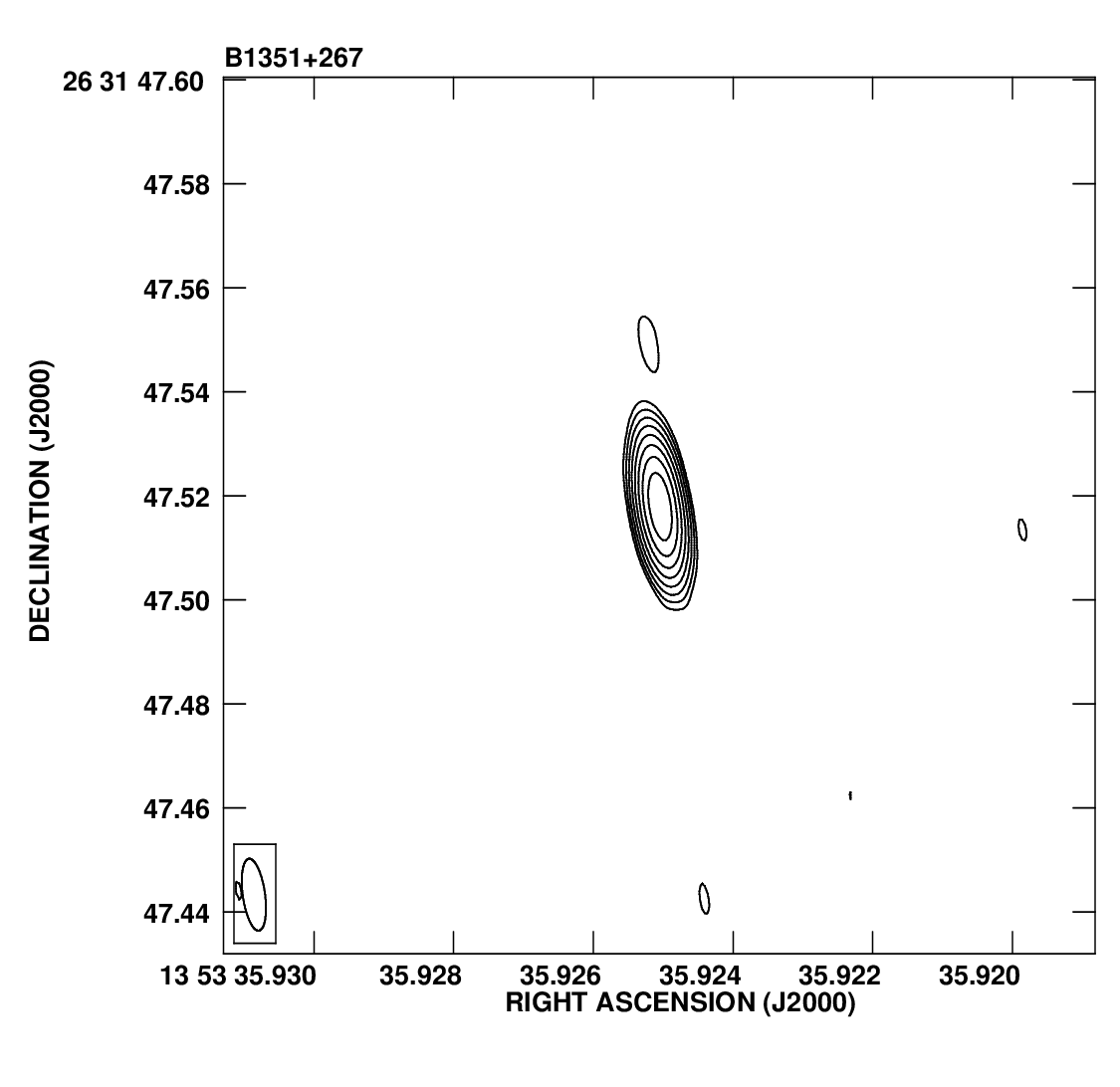}}
\caption{EVN image of the core of B1351+267. Contours are increased
by a factor of 2; the first contour level corresponds to 5\,$\sigma$ level 
which is 56\,$\mu$Jy/beam. The peak flux density is 13.1\,mJy/beam. The beam 
size is $13.9 \times 4.3$\,milliarcseconds at the position angle of 
8$\degr$.}
\label{fig:B1351_EVN}
\end{figure}

The EVN observations were carried out on 2009 November 9/10 at 1658\,MHz. 
The network consisted of 10 telescopes: Effelsberg, Lovell, Onsala, 
Medicina, Noto, Robledo, Shanghai, Toru\'n, Urumqi, and Westerbork. Two 
polarisations were recorded in eight 8-MHz sub-bands each. Because our targets 
were weak, phase referencing was used. The scans on targets and phase 
calibrators were 10- and 2-min. long, respectively and were placed as evenly 
as possible on the $u$--$v$ plane. We assigned 3.5\,hours of observing time
for \object{B0847+548}, 5\,hours for \object{J115139.6+335541}, and 4\,hours 
for \object{B1351+267}. The data reduction using AIPS was carried out in a 
standard way. Amplitude self-calibration was applied to the visibility data
before obtaining the final images; they are shown in Figs.~\ref{fig:B0847_EVN}, 
\ref{fig:J1151_EVN}, and \ref{fig:B1351_EVN}.
The noise in these images is typically 10\,$\mu$Jy/beam.

\begin{table}[b]
\caption{Observed parameters of the milliarcsecond structures}
\label{tab:obs}
\centering
\begin{tabular}{l r r c}
\hline
\hline
Object name & \multicolumn{2}{c}{Flux density} & Brightness\\
& Total\tablefootmark{a} & Fitted\tablefootmark{b} &
temperature\tablefootmark{c}\\
& [mJy] & [mJy] & [K]\\
\hline
B0847+548 & 10.0 & 8.3 & 6.65$\times 10^8$\\
J115139.6+335541 & 4.2 & 3.8 & 1.55$\times 10^8$\\
B1351+267 & 13.7 & 13.7 & 7.73$\times 10^9$\\
\hline
\end{tabular}
\tablefoot{
\tablefoottext{a}{Measured with AIPS TVSTAT utility.}
\tablefoottext{b}{Fitted to the strongest feature.}
\tablefoottext{c}{Based on the flux density of the strongest feature.}
}
\end{table}

Flux densities of the milliarcsecond structures were measured with AIPS 
TVSTAT utility. We also fitted single Gaussians to the brightest features 
in the source centres. Based on these figures, we calculated the 
brightness temperatures of those features using formula with Eq. (1) from 
\citet{Ulvestad2005}. All the above data are given in Table~\ref{tab:obs}. 
As discussed in Sect.~\ref{sect:cores}, very high brightness temperatures 
are indicative of jet production, albeit these jets could be very weak. This 
agrees with the observed morphologies: neither of the three 
milliarcsecond structures are of core-jet type as in the case of the 
lobe-dominated RLQs -- see discussion in Sect.~\ref{sect:cores}.

\begin{table}[b]
\caption{BH masses, bolometric luminosities, and Eddington ratios}
\label{tab:sdss}
\centering
\begin{tabular}{l c c c}
\hline
\hline
Object name & $\log(M_{BH}/M_{\sun})$ & $\log(L_{bol})$ & $L_{bol}/L_{Edd}$\\
& & [log(erg/s)]\\
\hline
B0847+548 & 9.406 & 45.541 & 0.011\\
J115139.6+335541 & 9.383 & 46.272& 0.061 \\
\hline
\end{tabular}
\end{table}

\subsection{Optical domain}

Spectra of two quasars, B0847+548 and J115139.6+335541, are available in the
Sloan Digital Sky Survey (SDSS)\footnote{Website: {\tt http://www.sdss.org/}}.
Both are very typical for QSOs: the 
continua rise towards shorter wavelengths and they are featured by both 
broad and narrow emission lines. For B0847+548, the H$\beta$, and for 
J115139.6+335541 the Mg\,II line can be best used for a measurement of BH 
mass, see e.g. \citet{CN2010} for an up-to-date review. The BH masses and 
related quantities for 105,783 quasars in the SDSS Data Release\,7
\citep{SDSSDR7} quasar catalogue \citep{DR7QSO}
have been recently calculated by \citet{Shen2010}. We 
extracted the BH masses, bolometric luminosities, and Eddington ratios for 
B0847+548 and J115139.6+335541 from that catalogue and listed them in 
Table~\ref{tab:sdss}. At the time of writing, the results of 
\citet{Shen2010} had not been accepted for publication yet so independent 
calculations were carried out by Niko{\l}ajuk (private communication). His results 
were very close to those of \citet{Shen2010}. Unfortunately, no B1351+267 
spectrum was available in the literature (\citet{EH2004} quoted only the 
redshift for this object). Therefore, we were unable to give the BH mass and 
Eddington ratio for it.

\section{Discussion}

\subsection{Cores of RLQs vs. cores of RQQs}\label{sect:cores}

Radio structures of QSOs are highly influenced by orientation effects as 
predicted by the AGN unification scheme. As a result, depending on the angle 
between the line of sight and the jets, RLQs can be perceived either as 
core- (beamed) or lobe-dominated (unbeamed). Doppler beaming and boosting 
makes the first group ideal objects for VLBI studies like, e.g. the 
Caltech--Jodrell Bank Flat-spectrum (CJF) sample survey \citep{Taylor1996} 
that encompasses 293 sources. Following the predictions based on the unification 
scheme, the core-jet structure overwhelmingly dominates them. On the
other hand, cores of 25 lobe-dominated QSOs from 3CR-based complete sample
were comprehensively investigated by \citet{Hough2002}, who observed them in
one or more epochs between 1981 and 1997 with various VLBI arrays at
several frequencies between 5 and 22\,GHz. Later, these observations were 
supplemented with VLBA and High Sensitivity Array (HSA) observations 
\citep{Hough2008} so eventually the nuclei in all 25 objects were imaged. 
Cores in three of them were unresolved and the remaining 22 showed one-sided 
jets located on the same side of the compact core as the one-sided 
large-scale jets seen in the VLA images. Given that no counter-jets were 
observed, the authors estimated that these objects were all orientated within 
$\sim$70$\degr$ to the line of sight. Their results were consistent with 
orientation-dependent relativistic beaming effects and unification of core- 
and lobe-dominated QSOs. In the latter class, the presence of jets is a 
clear signature that the objects belonging to it are not only apparently RL 
because of the lobes' dominance, but are truly RL, i.e. the energy transport from 
the core to the lobes, which is the prime cause of the existence of radio 
structures, is ongoing.

By definition, RQQs are devoid of radio lobes because their nuclei either
do not produce jets, or the jets they produce are too weak to inflate 
lobes. This does not preclude, however, the existence of weak radio cores 
there. Indeed, they are observed and the question that immediately emerges 
is whether that radio emission is associated with starbursts
\citep[see e.g.][]{SA1991}
or with weak-jet producing central engines. \citet{BB1998} sought the answer
for this question by means of the VLBA observations of 12 RQQs, because the detection of 
milliarcsecond structure implies that the brightness temperatures of the 
emission is $T_B \ga 10^6$\,K, while for typical supernova remnants $T_B \la 
10^5$\,K. Although the structures that they managed to map out were only 
indicative of jets, the measurements of brightness temperatures provided
strong evidence for jet-producing central engines in eight sources. Further 
investigations of RQQ cores by means of the VLBA observations were carried 
out by \citet{Ulvestad2005}. They observed six RQQ cores and got images for 
five of them: four were dominated by unresolved component, while the fifth 
showed quite a rich structure that was labelled ``two-sided jet'' by 
the authors. Note that the structure had a steep spectrum. Being aware of that
and expecting that our sources can also bear this feature, we decided to 
make our observations at 1.6\,GHz to attain better signal-to-noise ratios
at the cost of lower resolution.

The brightness temperatures measured by \citet{Ulvestad2005} were of the 
order of 10$^8$ or greater, ruling out the thermal origin of the observed 
radiation. Thus, the authors confirmed the hypothesis of 
\citet{Miller1993} that radio cores of RQQs were fundamentally similar to 
those in RLQs, while the RQQ jets were simply less powerful. The more 
general conclusion that emerges from their work -- but also from 
\citet{Blundell1996, BB1998, Blundell2003} -- is that the term ``RQQ'' 
should not be understood in a verbatim manner, namely an RQQ is not 
necessarily a QSO that is completely ``silent'', but weak in radio domain 
\citep{Barvainis2005}. What is much more important instead is that central 
engines of RQQs do not produce jets strong enough to transport the energy to 
the lobes. This agrees with the hypothesis put forward by 
\citet{Ghisellini2004} that all black hole/accretion disc systems in AGNs 
can produce some kind of outflow or jet, but only in a minority of 
cases, i.e. in truly radio-loud objects, the jet is successfully launched and 
accelerated to relativistic speeds. In the majority of cases, the jet is 
``aborted'', yet it is responsible for a relatively weak radio emission.

Our findings are fully consistent with those by \citet{Ulvestad2005} both in 
terms of brightness temperatures and milliarcsecond-scale morphologies:
B1351+267 is point-like, J115139.6+335541 is nearly point-like with two 
protrusions on the eastern and western side, and B0847+548 has two-sided mini-jets.
All three have low radio flux densities and yet high brightness temperatures.
Altogether, the galactic nuclei we imaged with the EVN are characteristic
of RQQs. The major difference, though, is pertinent to the large-scale 
overall radio structures: while the objects observed by \citet{Ulvestad2005} 
are truly RQ, ours are classified as RL because of the presence 
of the lobes. Nevertheless, we claim that the cores of the three quasars we 
deal with here not only resemble those in RQQs, but belong to the RQQ 
category and so these 
quasars are observed after a recent transition between the RL and RQ
states. Their cores herald the onset of the RQ state, whereas the lobes,
while still rendering the RL label 
formally applicable to these quasars, are actually only remnants of 
the {\em former} RL state. On the other hand, we find that our quasars do 
not fit to the model by \citet{BK2007} because it requires mass accretion rates 
substantially higher than Eddington, which is not the case, see 
Table~\ref{tab:sdss}. (This model has also been criticised on theoretical 
grounds by \citet{LB2008}.)

\subsection{Quasar--microquasar analogy}

Although few XRBs are persistent X-ray sources, the majority are variable. 
The analysis of their light curves at different X-ray bands shows a clear 
correlation between the luminosity and hardness, which led to distinction of 
two basic states in XRBs: high/soft (HS) and low/hard (LH). The 1$-$10\,keV 
radiation observed in the HS state is best explained as a thermal emission 
form the accretion disc, while the 20$-$100\,keV radiation of the LH state, 
when the spectrum becomes non-thermal with a typical photon index 
$\Gamma\sim$\,1.7 and an exponential cutoff around 100\,keV, is a result of 
Comptonisation of softer photons in the accretion disc corona of high-energy 
electrons. More precisely, the LH state can be regarded as a follow-up of 
the HS state switch-off: when the density of soft seed photons is greatly 
reduced, the resultant Comptonised spectrum hardens, see e.g. \citet{MR2004} 
for a review. Three additional states: ``quiescent'', ``very high'' (VH), 
and ``intermediate'' have been all identified for \object{GX\,339$-$4} by 
\citet{Markert1973}, \citet{Miyamoto1991}, and \citet{MK1997}, respectively. 
At least some of these three can be observed in other XRBs. \citet{Esin1997} 
attributed these five states -- in order of increasing luminosity these 
were: the quiescent, LH, intermediate, HS, and VH -- to different accretion 
rates, although the VH state could not be unified within the framework of 
their model of accretion flows around BHs. Also, according to them, 
different accretion modes were at work for particular states: in quiescent, 
LH, and intermediate states, the radiatively inefficient advection-dominated 
accretion flow (ADAF) was present, while for HS and VH the ADAF (inner) zone 
disappeared and the radiatively efficient accretion disc extended down to 
the innermost stable circular orbit (ISCO) around the BH.

The correlation between the X-ray-emission-based states in XRBs and their 
radio properties is particularly strong: the radio emission is generated in 
LH and VH states, while it drops in the HS state below detectable levels, most 
likely because of the physical disappearance of jets, see \citet{Fender1999} for 
a brief summary. The parallel existence of a clear RL/RQ dichotomy in XRBs 
and AGNs immediately leads to attempts to unify XRBs and AGNs based on their 
radio properties, i.e. to identify specific states observed in XRBs with 
classes of AGNs \citep{Meier2001, Maccarone2003, Falcke2004}. The following 
picture emerges: Sgr\,A$^\star$, LINERs, FR\,I radio galaxies, and BL\,Lac 
objects remain in a state analogous to the LH state of XRBs, RQQs and 
Seyfert galaxies represent the HS state, whereas FR\,II radio galaxies and 
RLQs behave as if they were in the VH state. More recently, this scheme was 
advanced by \citet{WC2008} who showed that FR\,I and FR\,II 
radio galaxies could have different accretion modes and the accretion mode 
in the majority of FR\,I galaxies could be of ADAF type similar to that in 
the LH state in XRBs. On the other hand, many RL objects had both the most 
powerful known radio jets {\em and} optical properties typical for HS state 
so the VH state, which is the only state where these two features could be 
combined, should be attributed to them.

In the framework of the above paradigm, the objects we observed, which
are effectively RQQs based on the status of their nuclei, are in
a state analogous to the HS state 
but given the presence of decaying large-scale radio structures, they must 
have entered that state only recently. This immediately leads to a 
conjecture that migration of an AGN to the analog of the HS state may
be a part of a more 
general evolutionary process. Taking into account all the similarities 
between XRBs and AGNs, such an evolution -- which is sped up by the factor 
of the AGN-to-XRB BH mass ratio -- should possibly be followed {\em 
directly} in XRBs.

\subsection{XRBs and AGNs -- common evolution?}

A new light on the XRB/AGN unification was shed by \citet{Fender2004}, but 
see also a simplified version of that paper: \citet{Fender2005}, who 
constructed a semi-quantitative model of XRB evolution. Piecing together XRB 
observational data and placing them on the hardness--intensity diagram 
(HID), they found an evolutionary path of XRBs, sometimes called the 
``turtle head'' diagram -- see Fig.\,7 in \citet{Fender2004}. The XRB 
timeline starts at the quiescence stage, then moves nearly vertically during 
the LH state. When it leaves that state, it turns left towards the VH state. 
At some point, usually corresponding to the peak of the VH state, the photon 
index $\Gamma$ rapidly increases, producing an internal shock in the outflow 
and the jet properties change, most notably its velocity. The final, most 
powerful jet has the highest Lorentz factor, causing the propagation of an 
internal shock through the slower-moving outflow in front of it. Following 
that, an XRB enters the HS stage when no jet is produced although some XRBs, 
like GRS\,1915+105, make repeated excursions back to VH state \citep[see 
e.g.][and references therein]{Sikora2007}. After a number of 
``loops'' in the HID (if any), an XRB enters the HS state permanently and 
returns, via intermediate state, back to quiescence.

The above path can be followed directly for XRBs because they evolve quickly 
enough, but given the identity of both the accretion process 
\citep{McHardy2006} and the jet generation mechanism in microquasars and 
quasars \citep{Tuerler2004} one should expect that this path is common to 
AGNs, albeit it is impossible to be followed within a human lifetime 
because of the timescaling factor, which reaches several orders of magnitude. 
What could be done, however, is to produce a HID analog for AGNs and check 
on a statistical basis if it resembles the HID for XRBs. This task was 
accomplished by \citet{Koerding2006}. Because the blackbody temperature of the 
accretion disc scales with black hole mass as $M^{-{1/4}}$, the SEDs of 
accretion discs around SMBHs peak in the optical/UV domain, not the X-ray, and 
a HID for AGNs would not contain any information about the disc emission. 
Therefore, instead of a HID, the authors constructed a 
disc-fraction/luminosity diagram (DFLD) plotting $L_D + L_{PL}$ against 
$L_{PL}/(L_D+L_{PL})$ -- where $L_D$ denoted the disc luminosity and 
$L_{PL}$ the luminosity of the power-law component -- for a sample of 4963 
SDSS Data Release\,5 quasars. What they found was a striking similarity
between the 
DFLD for AGNs and DFLD for a simulated sample of XRBs. This shows that in 
general a large population of AGNs follow the same pattern as XRBs. 
Consequently, not only particular classes of AGNs could be assigned to 
different XRB states in a ``static'' manner \citep{Meier2001, Maccarone2003, 
Falcke2004}, but also signatures of transitions between those states could be 
looked for in AGNs. In particular, thanks to the long decay times of the 
radio lobes \citep{KG1994}, a recent transition from RL VH state to RQ HS
state 
should be attributed to the objects with already radio-quiet cores and still 
observable relic radio lobes. The existence of these objects has been 
predicted by \citet{Fender2005}. We claim that the quasars with asymmetric 
overall radio structures we presented here fit those predictions.

Finally, it is worth mentioning at this point that our attempt to identify a 
transition in accretion properties of AGNs with a respective transition 
between spectral states of XRBs is not the first endeavour of this kind.
\citet{Marchesini2004} suggested that the transition of accretion properties 
among different types of RL AGNs alone could also have its analog in the XRB 
domain.

\section{Summary and final conclusion}

The unification of XRBs and AGNs has recently gained a substantial boost. 
A major step towards a full unification between these two classes were made 
by \citet{Fender2004}, who showed an evolutionary sequence for XRBs in the 
HID and by \citet{Koerding2006} who constructed a DFLD -- an analog of HID 
applicable for AGNs. One cannot directly show that AGNs are moving along an 
evolutionary track on a DFLD but, given that 
many characteristics of XRBs and AGNs have been found to be either analogous or 
identical, is it very reasonable to expect that the evolutionary sequence 
traced for XRBs should also be valid for AGNs. In this work, we showed the 
presence of transitions from the RL to the RQ state in three quasars, using their
large-scale radio structures as a historical record of the former RL state
and the micro-scale radio structures as an evidence of the present RQ state.
If, in line 
with \citet{Nipoti2005}, the XRB/AGN analogy indeed holds also for the 
evolution of these two classes of object, the quasars we investigated could 
be regarded as objects after a recent transition from VH to HS state.

The RL-to-RQ transition scenario outlined here has yet another consequence. 
According to the unification scheme \citep{Antonucci1993, UP1995}, the RL/RQ 
dichotomy should be attributed to the presence or absence of BH spin,
respectively. While 
small BH spins can still be the reason why some AGNs are RQ, according to 
some studies \citep[][and references therein]{Sikora2007}, BHs in {\em both} 
RLQs and RQQs may have large spins. For example, according to
\citet{Elvis2002} most SMBHs rotate rapidly. The same conclusion has been 
reached by \citet{Wang2006} based on an investigation of 12,698 quasars from SDSS.
The BHs in the three quasars we deal with 
here must have large spins to make it possible for these objects to have been RL in 
the past. They have become RQ now, but it is hard to find a mechanism forcing 
their BHs to spin down during such a transition. Eventually, when their 
lobes fade out completely, they will turn out to be {\em bona fide} RQQs, 
yet their BHs will still have large spins contrary to the paradigm.

\begin{acknowledgements}

\item AM is very grateful to Marek Sikora and also to Bo\.zena Czerny for 
inspiring discussions and particularly to Marek Niko{\l}ajuk for his help 
with the interpretation of the optical spectra and independent calculation 
of SMBH masses in the objects we investigated.

\item The European VLBI Network is a joint facility of European, Chinese, 
South African and other radio astronomy institutes funded by their national 
research councils.

\item This research has made use of the NASA/IPAC Extragalactic Database 
(NED), which is operated by the Jet Propulsion Laboratory, California 
Institute of Technology, under contract with the National Aeronautics and 
Space Administration.

\item Funding for the SDSS and SDSS-II has been provided by the Alfred P. 
Sloan Foundation, the Participating Institutions, the National Science 
Foundation, the U.S. Department of Energy, the National Aeronautics and 
Space Administration, the Japanese Monbukagakusho, the Max Planck Society, 
and the Higher Education Funding Council for England. The SDSS Web Site is 
http://www.sdss.org/. The SDSS is managed by the Astrophysical Research 
Consortium for the Participating Institutions. The Participating 
Institutions are the American Museum of Natural History, Astrophysical 
Institute Potsdam, University of Basel, University of Cambridge, Case 
Western Reserve University, University of Chicago, Drexel University, 
Fermilab, the Institute for Advanced Study, the Japan Participation Group, 
Johns Hopkins University, the Joint Institute for Nuclear Astrophysics, the 
Kavli Institute for Particle Astrophysics and Cosmology, the Korean 
Scientist Group, the Chinese Academy of Sciences (LAMOST), Los Alamos 
National Laboratory, the Max-Planck-Institute for Astronomy (MPIA), the 
Max-Planck-Institute for Astrophysics (MPA), New Mexico State University, 
Ohio State University, University of Pittsburgh, University of Portsmouth, 
Princeton University, the United States Naval Observatory, and the 
University of Washington.

\end{acknowledgements}

\end{document}